\documentclass[fleqn,twoside]{article}
\usepackage{espcrc2}

\usepackage{graphicx}
\usepackage{epsfig}
\usepackage[figuresright]{rotating}

\newcommand{\AmS}{{\protect\the\textfont2
  A\kern-.1667em\lower.5ex\hbox{M}\kern-.125emS}}

\title{Limits on the neutrino magnetic moment from the MUNU experiment}

\author{The MUNU collaboration: 
Z.~Daraktchieva \address[NE] {Institut de physique, A.-L. Breguet 1, 
                   CH-2000 Neuch\^atel, Switzerland},
J.~Lamblin\address[GR]{Laboratoire de Physique Subatomique et de 
Cosmologie, IN2P3/CNRS-UJF, 53 Avenue des Martyrs, F-38026 Grenoble, France},
O.~Link \address[ZH]  {Physik Institut, Winterthurerstr. 190, 
CH-8057 Zurich, Switzerland},
C.~Amsler \addressmark[ZH],
M.~Avenier \addressmark[GR] , 
C.~Broggini \address[PD]{INFN, Via Marzolo 8, I-35131 Padova, Italy},
J.~Busto \addressmark[NE],
C.~Cerna \addressmark[PD],
G.~Gervasio \addressmark[NE],
P.~Jeanneret  \addressmark[NE],
G.~Jonkmans \addressmark[NE],
D.H.~Koang \addressmark[GR],
D.~Lebrun \addressmark[GR],
F.~Ould-Saada \addressmark[ZH],
G.~Puglierin \addressmark[PD],
A.~Stutz \addressmark[GR],
A.~Tadsen \addressmark[PD],
J.-L.~Vuilleumier \addressmark[NE] 
}

\begin{document}\begin{abstract}
The MUNU experiment was carried out at the Bugey nuclear power
reactor. The aim was the study of
$\overline{\nu}_ee^-$ elastic scattering at low energy. The recoil
electrons were recorded in a gas time projection chamber, immersed in
a tank filled with liquid scintillator serving as veto detector,
suppressing in particular Compton electrons. The measured electron
recoil spectrum is presented. Upper limits on the neutrino
magnetic moment were derived and are discussed.
\vspace{1pc}
\vspace{0.5cm}
\end{abstract}

\maketitle

\section{Introduction}

The MUNU experiment was designed to study $\overline{\nu}_ee^{-}$
scattering at low energy, probing in particular the magnetic moment of 
the neutrino. The detector was set up near a nuclear 
power reactor in Bugey (France) serving as antineutrino source. We report here
on an analysis of data corresponding to 66.6 days of live time
reactor-on and 16.7 days reactor-off. 
 
\section{Magnetic moments}

Many experiments now show that neutrinos 
have masses, and that the weak eigenstates $\nu_{i}$ are superpositions of 
the mass eigenstates $\nu_{\ell}$:
\begin{equation}
         \nu_{\ell}=\sum_{i}U_{\ell i}\nu_{i}\;\;\;\;i=1,2,3;\;\ell=e,\mu,\tau.
\end{equation}
Intense  work is being done to determine more precisely the mixings
$U_{\ell i}$ and the masses $m_{i}$.

Besides masses, neutrinos can have magnetic moments. As shown in ref.
\cite{Kay82,Schr82,BeaV99}, the fundamental magnetic moments are
associated with the mass eigenstates, in the basis of which they are
represented by a matrix $\mu_{jk}$ ($j,k=1,2,3$). Dirac neutrinos can
have both diagonal and off-diagonal (transition) moments, while off
diagonal moments only are possible with Majorana
neutrinos. 

Astrophysical considerations put strong constraints on
Dirac neutrinos, and less strong ones on Majorana neutrinos (see
\cite{Raf96,MUNU1}). Upper limits applying in both cases
from stellar cooling of order 
$<10^{-12}\;\mu_{B}$ have been derived. Magnetic moments large enough would 
lead to spin-flavor precession in the toroidal magnetic field in the interior
of the sun, as discussed for instance in \cite{AkhL93,AkhP02}, further 
complicating the oscillation pattern of solar neutrinos resulting from 
masses and mixings.
 
These astrophysical bounds on magnetic moments, however, depend on
various assumptions and, to a certain extent, are model dependent.
Direct measurements are under much better control. So far the
best limits stem from experiments studying $\nu_ee^-$ or
$\overline{\nu}_ee^-$ scattering, and looking for deviations of the
measured cross-section from the one expected with weak interaction
alone. The cross-section is given by
\begin{eqnarray}
\label{eq:dsdt}
\lefteqn{\frac{d\sigma}{dT}  = } \nonumber\\
   & & \frac{G_{F}^{2}m_{e}}{2\pi}   \left[ (g_{V}+g_{A})^{2} 
   + (g_{V}-g_{A})^{2}\left(1-\frac{T}{E_{\nu}}\right)^{2} \right.  \nonumber  \\
   & &+  \left. (g_{A}^{2}-g_{V}^{2})\frac{m_{e}T}{E_{\nu}^{2}} \right] \nonumber\\
   & & \mbox{} +\frac{\pi \alpha^{2}\mu_{e}^{2}}{m_{e}^{2}}\frac{1-T/E_{\nu}}
           {T} 
\end{eqnarray}
with the contribution of the weak interaction in the first two lines,
and that from the magnetic moment $\mu_{e}$ in the last one
\cite{VogE89}. Here $E_{\nu}$ is the incident neutrino energy, $T$ the
electron recoil energy, and the couplings are given by
\[   g_{V}=2sin^{2}\theta_{W}+\frac{1}{2},\;\;
        g_{A}=\left\{ \begin{array}{rl}
           \frac{1}{2} & \mbox{ for $\nu_{e}$}\\
          -\frac{1}{2} & \mbox{ for $\overline{\nu}_{e}.$}
                      \end{array} \right.
\]
The relative contribution of the magnetic moment term increases with
decreasing neutrino and electron energies. Therefore it is essential
to have a low electron detection threshold, looking for neutrinos from
a low energy source. So far the sun and nuclear reactors have been
used.  Future experiments with radioactive sources are planned.

The measured squared magnetic moment $\mu_{e}^{2}$ depends on 
the mixings and, in case of a large distance $L$ from the source
to the detector, on 
the propagation properties of neutrinos. For vacuum 
oscillations it is given by:
\begin{equation}
(\mu_e)^2 = \sum_j\left|\sum_k
U_{ek}e^{ip_kL}\mu_{jk}\right|^2
\label{mudef}
\end{equation}
with
\[
p_k\cong
E_{\nu}-\frac{m_{k}^{2}}{2E_{\nu}} 
\]
the momentum of $\nu_k$ with mass $m_{k}$, for a given
neutrino energy $E_{\nu}$. In the case of matter
enhanced oscillations the dependence of the propagation eigenstates
on the local density must be taken into account \cite{BeaV99}.

\subsection{Solar neutrinos}

Here the source to detector distance is long, oscillations are relevant  
and the measured quantity is denoted
$\mu_e^{sol}$.  Super-Kamiokande has
measured with a very high statistical accuracy the electron recoil
spectrum from the scattering of solar $^8$B neutrinos
\cite{Kam99}. The total rate is reduced because of
neutrino oscillations, but the shape is seen to be in good agreement, 
within statistics, with that expected assuming weak interaction alone.
Beacom and Vogel \cite{BeaV99} looked to what extent excess counts at
the low energy end from a magnetic moment can be ruled out. The
assumption is made that such an excess is not compensated by a distortion
due to oscillations.  The limit $\mu_e^{sol}<1.5\cdot
10^{-10}$~$\mu_B$ at 90 \% confidence level (CL) was derived.

\subsection{Reactor neutrinos}

Nuclear reactors are strong sources of $\overline{\nu}_e$ with
energies ranging up to about 8 MeV. These are essentially produced in
the beta decay of fission fragments, with however a significant
contribution at low energy from nuclei activated by neutrons. Above
1.5-2.0 MeV the integral beta spectrum of the fission fragments of the
isotopes in which fission is predominantly induced by thermal neutrons
($^{235}$U, $^{239}$Pu, $^{241}$Pu) was measured. These isotopes
dominate. A simple procedure leads to the corresponding neutrino
spectra, which are thus known well, with a precision of order 5 \% or
better \cite{Zac86}. In $^{238}$U, fission is induced by fast
neutrons, and no such measurements are available. Less precise
calculations ($\sim\pm$10 \%) only can be used.  But this isotope
contributes only about 6-7 \% in a conventional reactor. The sum
spectrum above 2 MeV can be reconstructed with a precision of order
$\pm$5 \%, knowing the relative contributions of the various fissile
isotopes at a given reactor. Studies of $\overline{\nu}_ep
\longrightarrow e^{+}n$ scattering at reactors \cite{Zac86,Bug123}, which
sample the antineutrino spectrum above 2.5 MeV, 
show good agreement with these predictions. 

Below 1.5-2 MeV the neutrino spectrum can only be reconstructed from
inherently less precise calculations as shown in ref. \cite{Kop97}.
In that work it is mentioned that neutron activations of the
fissile isotopes, leading in particular to $^{239}$U, $^{239}$Np
and $^{237}$U, and of the fission products, contribute. The yields
have been estimated.

In reactor experiments the source to detector distance is short compared
to oscillation lengths, and the magnetic moment searched for
$\mu_e^{short}$ is given by eq. (\ref{mudef}) with L set to zero.

The Irvine group was the first to observe $\overline{\nu}_ee^-$
scattering \cite{ReGu76}.  The experiment was performed at the
Savannah River reactor, with a 16 kg plastic scintillation counter
surrounded by a NaI veto at 11 m from the reactor core.  A reactor-on
minus reactor-off signal was seen, with a threshold at 1.5 MeV.
Analyzing the electron recoil data with the most recent knowledge of
the Weinberg angle and the reactor spectrum, Vogel and Engel \cite{VogE89} 
found a
slight excess of events which can be explained by a magnetic moment 
$\mu_e^{short}$ of
order $(2-4)\cdot 10^{-10}$~$\mu_B$.

Also $\overline{\nu}_ee^-$ scattering was studied at the
Rovno reactor by a group from Saint Petersburg. The detector consisted
of a stack of silicon sensors, with a total mass of 75 kg. A reactor-on 
minus reactor-off signal was seen above 600 keV, with a signal to
background ratio of order 1:100. The limit $\mu_e^{short}<1.9(1.5)\cdot
10^{-10}$~$\mu_B$ at 95(68) \% CL was reported \cite{DerC93}. 

More recently the TEXONO collaboration installed an Ultra Low
Background High Purity Germanium detector, with a fiducial mass of
1.06 kg, near the Kuo-Sheng reactor in Taiwan \cite{TEXONO}. 
Here the approach is somewhat different.  The threshold on the 
electron recoil (12 keV) is extremely low. The reactor-on and reactor-off 
spectra were found to be identical within statistical errors. 
From that the limit
$\mu_e^{short}<1.3(1.0)\cdot 10^{-10}$~$\mu_B$ at 90(68) \% CL was
derived.

\section{MUNU}

In the aforementioned experiments the energy of the recoil electron
candidate only was measured. The MUNU collaboration \cite{MUNU1,MUNU2}
has built a detector of a different kind, in which the topology of
events is recorded. This allows a better event selection, leading to
a lower background. Moreover, in addition to the energy, 
the initial direction of an electron track can be measured. A second 
parameter, the electron
scattering angle, can therefore be reconstructed. This
allows to look for a reactor signal by comparing forward
electrons, having as reference the reactor to detector axis, with the
backward ones. The background is measured on-line, which
eliminates problems from detector instabilities, as well as from
a possible time dependence of the background itself.

\begin{figure}[htb]
\begin{center}
\hspace*{-0.7cm}
\epsfig{file=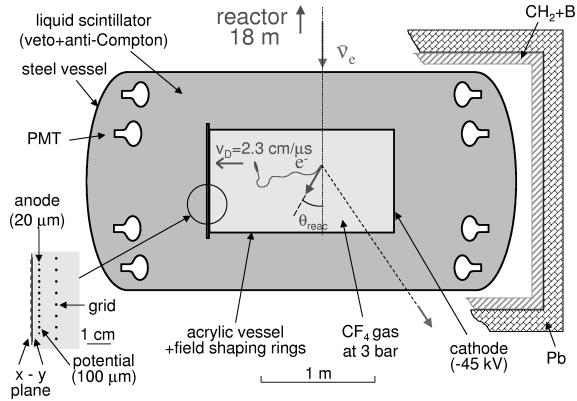,width=7.5cm}
\caption{The MUNU detector at the Bugey reactor.}  
\label{fi:MUNU0}
\end{center}
\end{figure}

\subsection{The detector}

The detector is described in details in ref. \cite{MUNU2}, and we 
only present here the essential features.
The detector, made from radiochemically clean materials, 
was installed at 18 m from the core of a commercial reactor in Bugey (France)
with a power of 2750 MWth. It emits neutrinos from the fission
fragments of $^{235}$U (54 \% on average over an annual reactor cycle),
$^{239}$Pu (33 \%), $^{241}$Pu (6 \%), and  $^{238}$U (7 \%).  
 
The central component of the MUNU detector consists of a time
projection chamber (TPC) filled with 3 bar of CF$_4$ gas (figure
\ref{fi:MUNU0}), acting as target and detector medium for the recoil
electron.  CF$_4$ was chosen because of its relatively low Z, leading
to reduced multiple scattering, its absence of protons, which
eliminates backgrounds from $\overline{\nu}_ep \longrightarrow e^+ n$
scattering, its good drifting properties and its high density. As
shown in figure \ref{fi:MUNU0} the gas is contained in a cylindrical
acrylic vessel of 1 m$^3$ volume (90 cm in diameter, 162 cm long,
total CF$_4$ mass 11.4 kg), which is entirely active. The drift field
parallel to the TPC axis (z-axis) is defined by a cathode on one side,
a grid on the other side, and field shaping rings at successive
potentials outside the acrylic vessel. The voltages were set to
provide a homogeneous field leading to a drift velocity of 2.14
cm$\cdot \mu$s$^{-1}$. An anode plane with 20 $\mu$m wires and a pitch
of 4.95 mm, separated by 100 $\mu$m potential wires, is placed behind
the grid, to amplify the ionization charge.  The integrated anode
signal gives the total energy deposit.

A pick-up plane with perpendicular x and y strips (pitch 3.5 mm)
behind the anode provides the spatial information in the x-y plane
perpendicular to the z-axis. The spatial information along the z-axis
is obtained from the time evolution of the signal. To reduce 
systematics when comparing forward and backward events, the TPC was
positioned orthogonally to the reactor-detector axis, which moreover
coincides with the bisecting line between x and y strips on the
pick-up plane. The anode wires are rotated by 45$^0$ with respect to
the x-y plane. The TPC is thus absolutely symmetric between backward
and forward directions with regards to the reactor-detector axis.

\begin{figure}[htb]
\begin{center}
\hspace*{-0.cm}
\epsfig{file=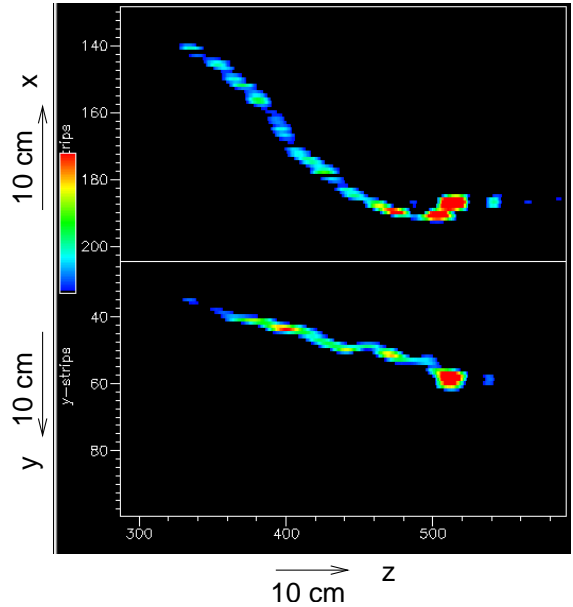,width=7.5cm}
\caption{A 1 MeV electron in the TPC. The x-z and y-z projections are
shown. The increased ionization which allows to determine the end of a
track is clearly visible.}
\vspace*{-1.cm} 
\label{fi:MUNU_track}
\end{center}
\end{figure}

The imaging capability of the TPC is illustrated in figure 
\ref{fi:MUNU_track} which shows an electron track.
The energy resolution and calibration is obtained by comparing spectra
measured using various $\gamma$ sources ($^{137}$Cs (662 keV), $^{54}$Mn 
(835 keV) and $^{22}$Na (1274 keV))
with simulations. The relative energy resolution is found to be 8 \%
(1 $\sigma$) at 1 MeV, correcting for small variations across the
anode plane. It scales with the power 0.7 of the energy, rather than with 
the square root. We think that this is because the electron attachement
in CF$_4$ in the strong field around the anode wires is rather high, 
affecting the statistics in the avalanche \cite{MUNU2,Lam02}. The gain
stability is monitored regularly throughout the data taking period,
with sources and by measuring the spectrum of cosmic muons crossing
the TPC. The data are corrected for small instabilities.  The angular
resolution is around 10$^0$ (1 $\sigma$) at 1 MeV, for tracks scanned
visually, as derived from Monte-Carlo simulations, with a slight
angular dependence \cite{Dar03}. The energy dependence is only weak 
above 700 keV.

The acrylic vessel is immersed in a steel tank filled with liquid
scintillator and viewed by photomultipliers, 24 on each side. 
It acts as a veto counter
against the cosmics and as an anti-Compton detector to reduce the
background from $\gamma$'s. The scintillator also sees 
the light produced by the
avalanche around the anode wire,
providing an additional measurement of the ionisation charge. The
primary light of heavily ionizing particles such as $\alpha$'s is seen
as well. A few $\alpha$'s are observed from remaining surface
contaminations on the
cathode. The primary light of minimum ionizing particles confined in
the gas volume however is below detection 
threshold.

The TPC threshold is set at 300 keV, and that in the scintillator at
100 keV, with a minimum of 5 photomultipliers hit. We note that the
singles rates in the photomultipliers show no azimuthal dependence,
which could result from hot spots.  

In normal data taking TPC events
above threshold are read out, provided they are not in coincidence
with a 200 $\mu$s signal started by scintillator pulses above 22
MeV. This eliminates direct cosmic hits, or neutrons associated with
them. The 48 photomultipliers are read out first. The light from the 
photomultipliers on the anode side and the cathode side is 
compared. Reading proceeds only if the relative difference is less than 
$\pm$30 \%.
Events corresponding to real tracks inside the gas volume were found
to always fall within these limits \cite{Lin03}. Discharges however can result 
in a larger imbalance, in which case they are readily eliminated.
This helps reducing the average readout time.

\section{Event selection}

Good events are single electrons contained in the TPC volume.  The
selection of neutrino scattering events proceeds in two steps. First
an automatic filtering eliminates obviously bad events, namely events
\begin{itemize}
\item {identified as $\alpha$'s or discharges from their topology
(high ionization in a small volume),}
\item {or in delayed coincidence (80 $\mu$s, corresponding to the TPC
length) with a signal in the scintillator,}
\item {or events not contained in a fiducial volume of 42 cm radius,}
\item {and finally events with a fast rise time, due to particles
crossing the amplification gap between the pick up plane 
plane and the grid, in either direction. Here the avalanche light signal 
is used, as described in ref. \cite{MUNU2}.}
\end{itemize}
The precise live time is derived in this process on a daily basis. It
is found to fluctuate around 65 \%.  It is limited primarily by the
total veto time of the scintillator (11 \%) and the dead time of the
TPC itself, caused by the relatively long data read-out and data transfer
time (24 \%).

Then a final scan is performed. In a separate publication, we reported
first results from a crude automatic procedure \cite{MUNU3}, carried
out with a pattern recognition program.  Here we present an analysis
based on a visual scan of events, which has the advantage of cleaner
event selection. Also the risk of misidentifying the beginning of a
track is much reduced. This approach is time consuming, however.  To
minimize the work load, that analysis was restricted to energies above
700 keV. Results from an improved version of the automatic scanning
procedure will be discussed in a subsequent section.  The data sets
corresponds to 66.6 days of live time reactor-on and 16.7 days
reactor-off.
\begin{figure}[hbt]
\begin{center}
\epsfig{file=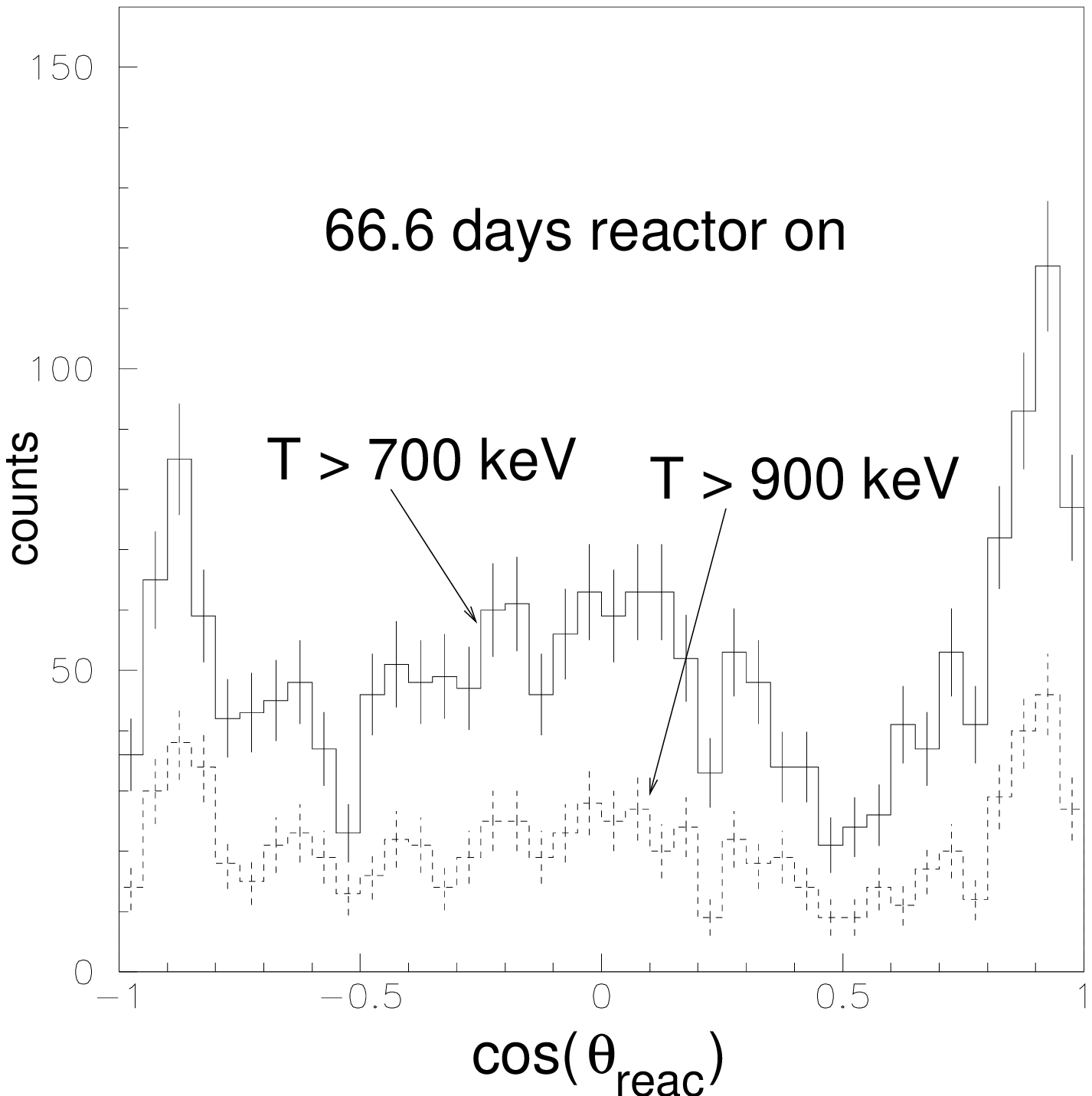, height=6.cm}
\vspace*{-1.0cm}
\epsfig{file=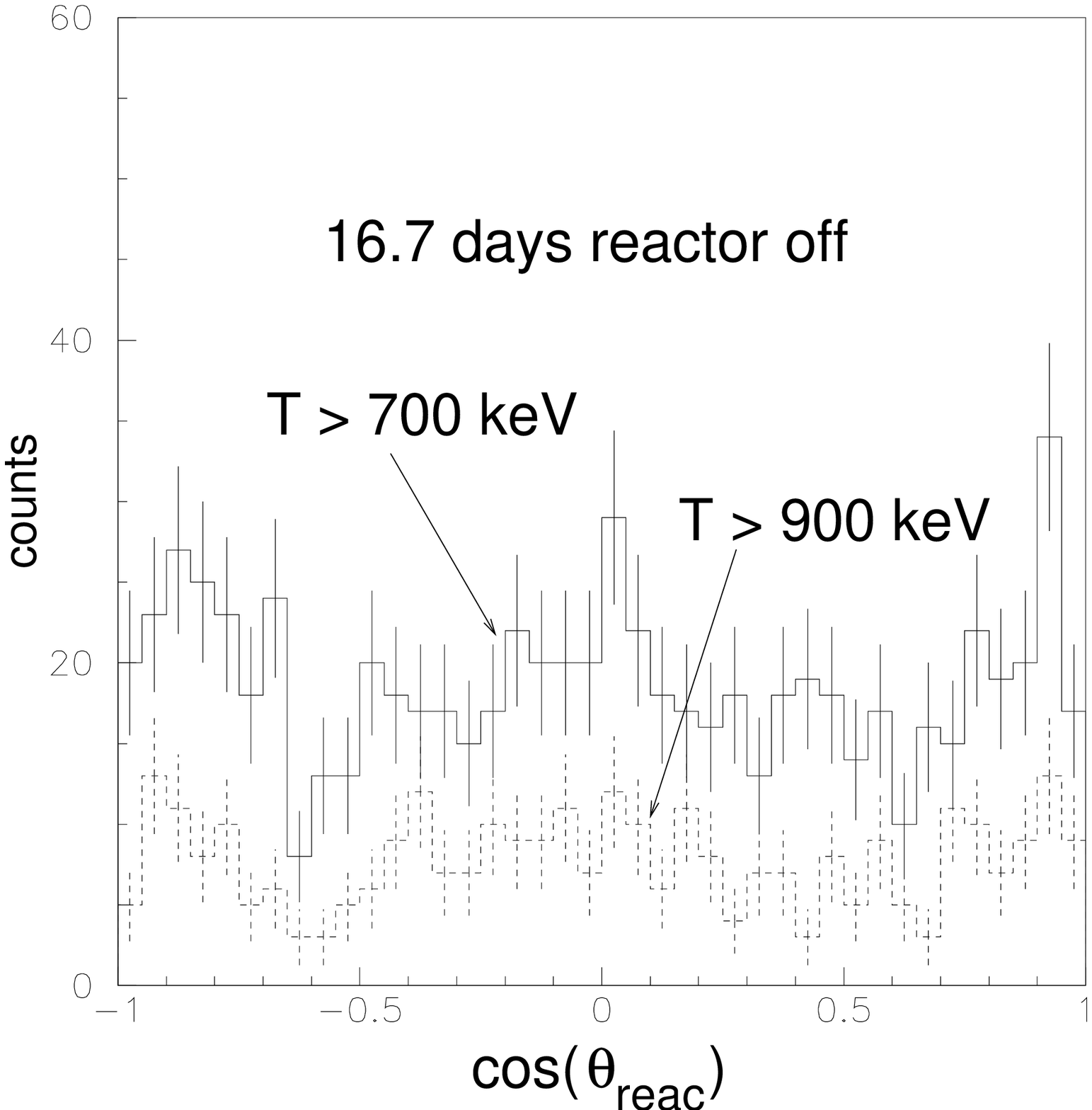, height=6.cm}
\caption
{Distribution of $\theta_{reac}$ for single electron events, 
top reactor-on, bottom reactor-off, visual scanning.}  
\label{fi:theta_ron}
\end{center}
\end{figure}

In this second scan both the $x-z$ and $y-z$ projections of an event
are scrutinized, as well as the evolution in time of the anode and
scintillator signals.  Continuous electron tracks only are
retained. The end of the track is identified from the increased energy
deposition, due to the higher stopping power, as exemplified in figure
\ref{fi:MUNU_track}. Events with a second high deposition along the
track, in particular near the other end, are discarded. This improves
the background suppression, at the cost of events with a delta
electron, which however contribute negligibly.

Then the tangent at the start of the track is determined by eye. From
that the angles $\theta_{reac}$ with respect to the reactor
core-detector axis is determined, as well as the angle $\theta_{det}$
with respect to the TPC axis.

An excess of events from the anode side was observed. It is presumably
due to additional activities, resulting from the greater complexity of
the readout system,
and to a larger inactive volume in the scintillator because of the stronger
and thicker acrylic lid. For that reason only electrons emitted in the half
sphere from the cathode side ($\theta_{det}<90^0$) are accepted. This reduces 
the acceptance by a factor 2, but leads to a better signal to background 
ratio.

\section{Results from the visual scan}

In figure
\ref{fi:theta_ron} we show the distribution of
$\cos(\theta_{reac})$ of single contained electrons, for both reactor-on 
and reactor-off. The slightly non linear angular response of the
TPC, and its geometry, explain the accumulation of events at
$\cos(\theta_{reac})$ around 1, -1 and 0 for reactor-off. The
distribution is however identical in forward
($\cos(\theta_{reac})\simeq 1$) and backward
($\cos(\theta_{reac})\simeq -1$) directions. 
The reactor-on spectrum shows a
clear excess of events in forward direction from $\overline{\nu}_ee^-$
scattering. Uncertainties from instabilities of the gain, the veto
rate, the live time, cancel out in this forward minus backward
comparison.

For each electron event the neutrino energy $E_{\nu}$ is reconstructed
from the electron recoil energy $T$ and from the scattering angle,
taken as $\theta_{reac}$. Forward event candidates are
defined as those with positive neutrino energy: $E_{\nu}>0$. To select
backward events the neutrino energy which must be positive is that
calculated using $\pi-\theta_{reac}$ as scattering angle. This
procedure has an acceptance close to 100 \% above 700 keV, as 
determined by Monte-Carlo simulations. It takes into account that electrons 
are emitted in an 
narrower cone whith increasing energies, and is somewhat 
more sophisticated  than the application of a crude cut on 
$\cos(\theta_{reac})$. In practice, however, with a threshold of 700 keV,
it produces almost the same result as the application of a cut
$\cos(\theta_{reac})>0.7$ or $\cos(\theta_{reac})<-0.7$.
  
\begin{figure}[htb]
\vspace*{-0.5cm}
\begin{center}
\epsfig{file=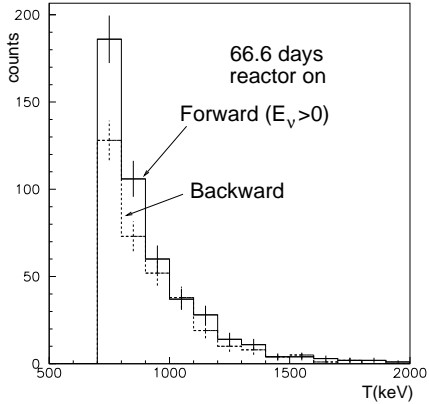, height=6.0cm}
\vspace*{-1.cm}
\caption{Energy distribution of forward and backward events, reactor-on, 
visual scanning.}  
\label{fi:fbe}
\end{center}
\end{figure}

The energy distributions of both forward and backward
events are displayed in figure \ref{fi:fbe}. A clear excess of forward events
(458 in total) over backward events (340) is seen. 
The total forward minus backward count rate above 700 keV is thus
1.77$\pm$0.42~day$^{-1}$.  The background, given by the backward
events, shows a steep low energy component ending at about 1.2
MeV. Above it is seen to be fairly low, and much flatter. The
background was observed to increase slightly during the course of the
experiment, by about 10 \% in 6 months, possibly because of outgassing. 
A small drop in
efficiency of the anti-Compton, even without an observable reduction
in count rate, could also explain this.

\begin{figure}[htb]
\vspace*{-0.5cm}
\begin{center}
\epsfig{file=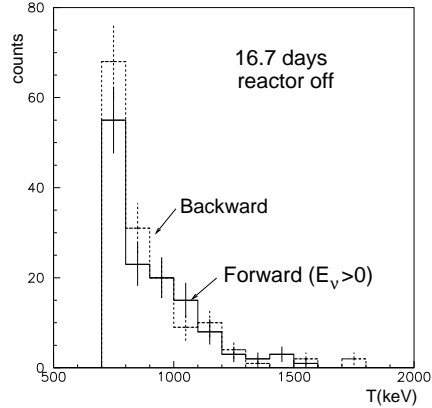, height=6.0cm}
\vspace*{-1.cm}
\caption{Energy distribution of forward and backward events, reactor-off, 
visual scanning.}  
\label{fi:fbeoff}
\end{center}
\end{figure}

As a cross check the reactor-off recoil spectra, 
both forward and backward, measured after the
reactor-on period, were reconstructed as well, and are displayed in
figure \ref{fi:fbeoff}. They are indeed identical within statistics,
the integrated forward minus backward rate above 700 keV being
$-1.02\pm1.00$~day$^{-1}$. 

\begin{figure}[htb]
\vspace*{-1.0cm}
\begin{center}
\epsfig{file=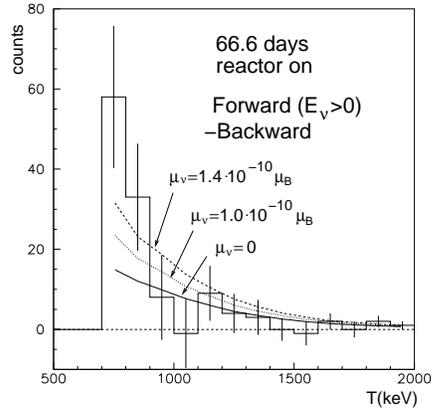, height=6.0cm}
\vspace*{-1.cm}
\caption{Energy distribution, forward minus backward events, reactor-on, 
visual scanning.}  
\label{fi:fminusbe}
\end{center}
\end{figure}

The difference of the forward minus backward reactor-on spectra is
shown in figure \ref{fi:fminusbe}.  The expected event rate was
calculated using the best knowledge of the reactor spectrum as
described above, and taking into account the known activations of
fissile isotopes and fission products. The uncertainty is around 5 \%
above 900 keV, as discussed previously, and larger below.  The various
acceptances of the event selection procedure were determined from
measurements with sources, and by Monte Carlo simulations using the
GEANT3 code. Data sets including recoil electrons from neutrino
scattering were produced, filtered and scanned just as the real
data. The containment efficiency in the 42 cm fiducial radius for
recoil electrons was found to vary from 63 \% at 700 keV, 50 \% at 1
MeV to 12 \% at 2 MeV. The relative uncertainty is of order 2-3
\%. Some tracks with weird topologies cannot be reconstructed,
reducing the acceptance by  4 \%, relatively
speaking. Similarly the remaining acceptances together, including that of the
$E_{\nu}>0$ cut, 
lead to an additional 6 \% reduction.
Neutrino interactions in the copper cathode give rise to electrons
which can escape into the gas volume. These cannot be vetoed. Taking into 
account all cuts they increase the expected rate between 700 keV to 2 MeV
by some 3 \%. The relative uncertainty on the
global acceptance is of order 7 \%, leading to a total uncertainty of
9 \% on the total expected rate.

The total expected rate above 700 keV assuming a vanishing magnetic moment
was found to be 1.02$\pm$0.1~day$^{-1}$, to be compared with
1.77$\pm$0.42~day$^{-1}$ measured, as mentioned above. There is thus a
certain excess of measured events, which has however only a small
statistical significance. The calculated energy distribution is shown
in figure \ref{fi:fminusbe}. The excess counts are seen to be in the
region below 900 keV.

To be more quantitative, a $\chi^2$ was calculated using 100 keV bins
as shown in figure \ref{fi:fminusbe} from 700 keV to 1400 keV, and then a
bin from 1400 to 2000 keV. Gaussian
statistics applies then to all bins.  The error on the expected rate
is small in comparison to the statistical uncertainties, and turns out
to be negligible. The $\chi^2$ was calculated this way for several values of
the squared magnetic moment $(\mu_e^{short})^2$, constraining it to the region 
$(\mu_e^{short})^2=(0.77\pm0.92)\cdot 10^{-20}$~$\mu_{B}^2$. This is 
consistent with a vanishing magnetic moment. Renormalizing to
the physical region $(\mu_e^{short})^2\ge 0$ we find the limit 
$\mu_e^{short}<1.4(1.2)\cdot
10^{-10}$~$\mu_{B}$ at 90(68)\% CL. 

Nevertheless even the best fit is not very satisfactory. The $\chi^2$
(9.9 for 7 degrees of freedom) is on the high side. More troublesome,
as seen in figure \ref{fi:fminusbe}, the inclusion of a magnetic
moment in the calculated spectrum improves the agreement with the data
in the two first bins from 700 to 900 keV, but makes it worse in the
upper ones.

These two first bins are solely responsible for the high total
rate. We note that such an excess is also visible, unfortunately also
with limited statistical precision, and appearing at somewhat smaller
energies, in the data of ref. \cite{DerC93}. This excess may well
result not from a magnetic moment, but instead from sources not taken
into account when evaluating 
the reactor neutrino spectrum. As mentioned above, the
low energy part, which contributes significantly to the electron
spectrum below 0.9 MeV, is not so well known. Additional neutron
activations may contribute, beyond the known ones of the fissile isotopes and
the fission products. In that sense it seems safe at this stage to
restrict the analysis to electron energies
above 900 keV. There the measured event rate is $0.41\pm
0.26$~day$^{-1}$, in good agreement with the expected one $0.62 \pm
0.05$~day$^{-1}$. From the event rate and the energy distribution,
using the same prescription as above, the allowed range
$(\mu_e^{short})^2=(-0.95\pm 0.95)\cdot 10^{-20}$~$\mu_{B}^2$ is
found.  The best $\chi^2$ is 1.51 for 5 degrees of freedom. This
yields the limit
\begin{equation}
\mu_e^{short}<1.0(0.8)\cdot 10^{-10}$~$\mu_{B}\mbox{~at~90(68)~\%CL}
\end{equation}  
somewhat more stringent than the previous
one.

\section{Automatic scanning} 

Extending the visual scan to energies below 700 keV is not feasible
because of the rapid increase of the background below that energy,
leading to an unmanageable workload. But the automatic scanning
procedure mentioned in \cite{MUNU3} has been improved, and used on the
same data set \cite{Lin03}. The program first identifies electron tracks,
searches the vertex, and then
fits the beginning of the track.

\begin{figure}[htb]
\vspace*{-1.0cm}
\begin{center}
\epsfig{file=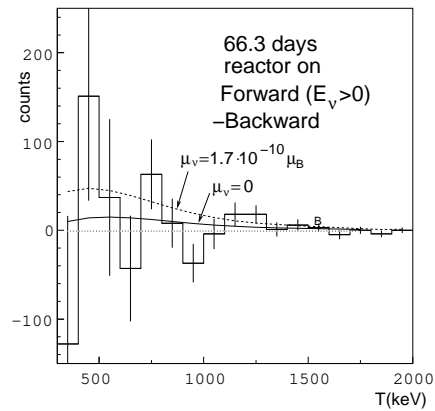, height=6.0cm}
\vspace*{-1.cm}
\caption{Energy distribution, forward minus backward events, reactor-on, 
automatic scanning.}  
\label{fi:f-be_auto}
\end{center}
\end{figure}

Electron tracks as identical as possible to real tracks were produced
by Monte-Carlo. The electronics noise was included. The tracks were
analyzed with the same program to determine the global acceptance, as
well as the angular resolution. The angular resolution is inferior to
that obtained in the visual scanning, varying from 31$^0$ at 300 keV,
to 18$^0$ at 700 keV and finally 12$^0$ at 2 MeV. This reduces the
acceptance of the $E_{\nu}>0$ cut in particular at low energy. This,
combined with poorer electron track identification, leads to an
acceptance of 20 \% at 300 keV, increasing to 54 \% at 1 MeV and
remaing constant above that energy. Events with $\theta_{det}<100^0$
were taken. The other cuts and acceptances are the
same as in the visual scanning.  The background suppression
is inferior by a factor 3 or more, so that the statistical precision
is less. But the method can be applied to energies extending down to
300 keV. We show in figure \ref{fi:f-be_auto} the forward minus
backward spectra for reactor-on. It is in good general agreement
with the spectra from the eye scan. The high bin at 750 keV is
reproduced.  But, admittedly with relatively large statistical errors,
a catastrophic rise when going to lower energies can be ruled out.
Applying the same procedure as above to the data one obtains the limit
$ \mu_e^{short}<1.7\cdot 10^{-10}$~$\mu_{B}$ at 90 \% CL.  It does
not change with the threshold in the range 300 to 800 keV, the better
sensitivity at low energy being offset by the larger statistical
uncertainties.

\section{Conclusions}

The MUNU experiment studied $\overline{\nu}_ee^-$ scattering at low energy
near a nuclear reactor. For electron recoil energies above 900 keV, good
agreement is seen with expectations assuming weak interaction alone. From
this the following limit on the magnetic moment of the neutrino can be derived:
$\mu_{e}^{short}<1.0(0.8)\cdot 10^{-10}$~$\mu_{B}$ at 90(68) \% CL, 
limited primarily by statistics. 

In any event this bound rules out the possible indication for a
relatively large magnetic moment from the Savannah River experiment
\cite{VogE89}, and improves on the limit from ref. \cite{DerC93} and
\cite{TEXONO}. It
is somewhat more stringent than the limit on $\mu_{e}^{sol}$ from solar
neutrinos \cite{BeaV99}, which however does not apply to exactly the same 
quantity.

The authors want to thank Edf-CNPE Bugey for the hospitality. This
work was supported by IN2P3, INFN, and SNF.

\end{document}